\documentclass[USenglish,twocolumn]{article}

\usepackage[utf8]{inputenc}
\usepackage[big]{dgruyter}
\usepackage{color}
\usepackage{multirow}

\newcommand{\bra}[1]{{\left\langle{#1}\right\vert}}
\newcommand{\ket}[1]{{\left\vert{#1}\right\rangle}}
   
\begin{document}

  \articletype{Research Article{\hfill}Open Access}

\author[1]{Rui Li}
\author[2]{Unai Alvarez-Rodriguez}
\author*[3]{Lucas Lamata}
\author[4]{Enrique Solano}

\affil[1]{Department of Physics, Zhejiang University, Hangzhou 310027, China, and Department of Physics and Research Center OPTIMAS, University of Kaiserslautern, 67663 Kaiserslautern, Germany, E-mail: lr6677@me.com}
\affil[2]{Department of Physical Chemistry, University of the Basque Country UPV/EHU, Apartado 644, 48080 Bilbao, Spain, E-mail: unaialvarezr@gmail.com}
\affil[3]{Department of Physical Chemistry, University of the Basque Country UPV/EHU, Apartado 644, 48080 Bilbao, Spain, E-mail: lucas.lamata@gmail.com}
\affil[4]{Department of Physical Chemistry, University of the Basque Country UPV/EHU, Apartado 644, 48080 Bilbao, Spain, and IKERBASQUE, Basque Foundation for Science, Maria Diaz de Haro 3, 48013 Bilbao, Spain,  E-mail: enr.solano@gmail.com}

  \title{\huge Approximate Quantum Adders with Genetic Algorithms: An IBM Quantum Experience}
  \runningtitle{An IBM Quantum Experience}

  \begin{abstract}
{It has been proven that quantum adders are forbidden by the laws of quantum mechanics. We analyze theoretical proposals for the implementation of approximate quantum adders and optimize them by means of genetic algorithms, improving previous protocols in terms of efficiency and fidelity. Furthermore, we experimentally realize a suitable approximate quantum adder with the cloud quantum computing facilities provided by IBM Quantum Experience. The development of approximate quantum adders enhances the toolbox of quantum information protocols, paving the way for novel applications in quantum technologies.}
\end{abstract}

  \journalname{Quantum Meas. Quantum Metrol.}

\DOI{https://doi.org/10.1515/qmetro-2017-0001}
  \startpage{1}
  \received{May 15, 2017}
  \accepted{June 23, 2017}

  \journalyear{2017}
  \journalvolume{4}
 
\maketitle
\section{Introduction}
Addition is arguably the most fundamental operation in mathematics, while adder machines are central to computation in general. The quantum adder was defined as a plausible quantum operation adding two unknown quantum states, encoded in different quantum systems, onto a single physical register~\cite{qa1,qa2}. This operation was proven to be forbidden by consistency relations involving a global phase in the description of the summands and the sum.  However, a deterministic approximate quantum adder was considered via the use of unitary techniques~\cite{qa1}. At the same time, a probabilistic quantum adder with partial prior knowledge of the summands was proposed~\cite{qa2} and has been realized in the lab~\cite{qa3,LaflammeExp}. In parallel, a study of quantum adders in the context of closed timelike curves has been developed~\cite{qa4}. The use of approximate quantum adders as constituents of quantum algorithms and protocols is certainly promising, as showcased in a recent result with the first application of a quantum adder~\cite{qa5}.

\begin{figure*}[t]
{\includegraphics[width=\textwidth]{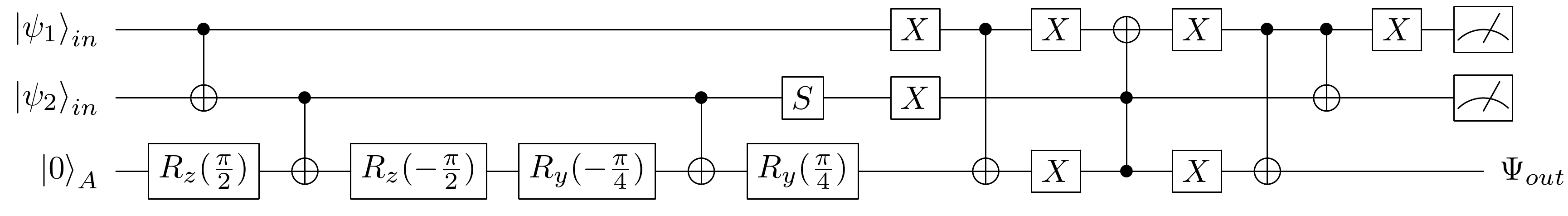}} 
\caption{{\bf Quantum circuit for the basis adder $U$.} }
\label{eq05}
\end{figure*}

The gate decomposition problem, which is often present in the design of quantum information experiments,  aims at finding the optimal quantum circuit that implements a given protocol. The minimization of the number of entangling gates is a crucial element, given that the experimental resources are limited. Although there are methods that simplify this task \cite{gd1,gd2,gd3,gd4,gd5}, there is no solution that provides an optimal decomposition for a general $n$-qubit protocol.  Genetic algorithms (GAs) include a set of optimization techniques inspired by natural selection, which is the key mechanism of evolution in biology. First considered by Alan Turing \cite{ga1} and refined in the following decades \cite{ga2,ga3,ga4,ga5}, the history of GAs is full of successful applications in science and technology. They have been found to be useful also in the context of quantum simulation and quantum information \cite{qga1,qga2,qga3,qga4,qga5}, as an alternative to different optimization techniques \cite{r1,r2}, among others. For instance, gate decomposition problems may be directly encoded as a sequence of instructions that conform the genetic code in the language of GAs. 

In this article, we propose a genetic algorithm optimization of approximate quantum adders~\cite{qa1}. Our work has been motivated by the search of an approximate quantum adder with a compromise between fidelity and the number of single-qubit and two-qubit gates required. In what follows, we will first discuss the global phase ambiguity preventing the existence of a general quantum adder machine~\cite{qa1} and how to overcome this problem. Then, we will explicitly provide the gate decomposition of the original approximate quantum adder for qubits~\cite{qa1} and analyze its feasibility. Subsequently, we will explain the use of genetic algorithms to find an optimal gate decomposition that yields an approximate quantum adder with average ideal fidelity of $95\%$. Moreover, when we reduce the allowed number of gates in this protocol, using parameters of a generic superconducting circuit platform, we still obtain an appreciable average fidelity with gate error estimation of $87\%$~\cite{superconducting}. Finally, we use the quantum computer of IBM Quantum Experience~\cite{IBMQE,Latorre} facility to experimentally realize our approximate quantum adder with genetic algorithms.

\section{Results}

\subsection{Self-consistent definition of a quantum adder.}
When proposing an approximate quantum adder, we first have to make it self-consistent with respect to the global phase variation. The latter does not affect a possible experimental realization but modifies the definition of the ideal output, and therefore the fidelity function. 

The absence of global phase invariance lies at the heart of the no-go theorem for a quantum adder~\cite{qa1}. There are two ways to fix this feature of a quantum adder machine. The first option is to modify the definition of the quantum adder by inserting a relative phase factor $e^{i\phi}$ to account for the ambiguity in the global phase of the initial state. Thus, instead of matching the output state of the quantum adder with  $\psi_1+\psi_2$, we match it with $\psi_1+e^{i\phi}\,\psi_2$ for a certain $\phi$, as originally proposed~\cite{qa1}. The second option is to restrict the domain of the quantum adder from the whole Hilbert space to a self-consistent region, and to fix the value of the relative phase to avoid phase ambiguity. The first approach would prevent us from exactly knowing the ideal state of reference after summation and, for certain inputs, we would not be able to distinguish the outcome states $\ket{0}$ and $\ket{1}$. We thus choose the second approach to circumvent the global phase problem without changing the most natural definition of a quantum adder, by restricting our two input states to take the form,
\begin{align}
\ket{\psi_i}_{in} &= \left(\begin{array}{c}
\cos\, \theta_i \\
\sin\,\theta_i  
\end{array}\right) .
\end{align} 
Here, $\theta_i$ goes from 0 to $\pi/2$. The ideal reference state after addition, with $1/N$ as the normalization factor, is
\begin{align}
\ket{\Psi_{id}}= \ket{\psi_1}_{in}+\ket{\psi_2}_{in}&=\frac{1}{N} \left(\begin{array}{c}
\cos\, \theta_1 + \cos\, \theta_2\\
\sin\,\theta_1 + \sin\,\theta_2 
\end{array}\right).
\end{align}
Notice that, by choosing this parametrization, we are effectively selecting the value of both external and internal phases $\phi=0$.

\subsection{The basis quantum adder.}

Suppose we want a quantum adder machine to add correctly the elements of the chosen computational basis. Then, the adding machine $U$ of the proposed basis quantum adder must have the following properties~\cite{qa1},
\begin{align}\label{eq01}
&U \ket{00}_S\ket{0}_A = \ket{B_1}\ket{0}, \hspace{0.25cm} U \ket{01}_S\ket{0}_A = \ket{B_2}\ket{+}, \\ \nonumber &U \ket{10}_S\ket{0}_A = \ket{B_3}\ket{+}, \hspace{0.25cm} U \ket{11}_S\ket{0}_A = \ket{B_1}\ket{1},
\end{align}
 where the subscripts $S$ and $A$ stand for system and ancillary qubits respectively, $B_i$ stand for the states of the two residual qubits to be discarded in the outputs, and $\ket{\pm}=\frac{1}{\sqrt{2}}(\ket{0}\pm\ket{1})$. To uniquely define our quantum adder, we need to complete the action of $U$ on the computational basis when the ancillary qubit is in state $\ket{1}$. We choose the definition of the basis quantum adder $U$ in the following manner,
\begin{align}
\nonumber &U \ket{000}=\ket{000}, \hspace{0.25cm} U \ket{010}=\ket{01+}, \hspace{0.25cm} U \ket{100}=\ket{10+},\\ \nonumber &U \ket{110}=\ket{001}, \hspace{0.25cm}  U \ket{001}=\ket{110}, \hspace{0.25cm} U \ket{011}=\ket{01-}, \\  &U \ket{101}=\ket{10-}, \hspace{0.25cm} U \ket{111}=\ket{111} ,
\end{align}
such that it can be decomposed as
\begin{equation}
 U=P^{(2,7)}~ U_{\rm CNOT}^{(1,2)}~ U_{\rm CHad}^{(2,3)} ~ U_{\rm CNOT}^{(1,2)} \, ,
 \end{equation}
with $P^{(2,7)} = U_{\rm CNOT}^{(\bar{1},2)}~U_{\rm CNOT}^{(\bar{1},3)}~ U_{\rm Toff}^{2\bar{3},1}~U_{	\rm CNOT}^{(\bar{1},3)}~U_{\rm CNOT}^{(\bar{1},2)}$. Here, $U_{\rm CNOT}^{(i,j)}$ stands for controlled-NOT (CNOT) gate with the $i$th qubit to be the control and the $j$th qubit to be the target, $U_{\rm Toff}^{ij,k}$ denotes the Toffoli gate, with qubits $i$ and $j$ controlling the $k$th one. Moreover, $U_{\rm CHad}$ is the controlled-Hadamard gate, and the overbar symbol on the control qubit means that the role of 0 and 1 levels is exchanged in this qubit. The whole protocol of the basis adder $U$ can be depicted with the quantum circuit in Fig.~\ref{eq05}.
    
There, $X$, $S$, and $R_{\alpha}(\theta)$ correspond respectively to the Pauli $X$ gate, the phase gate, and  rotations of $\theta$ in the $\alpha$ Pauli matrix. According to the principle of implicit measurement, any undetermined quantum wires (qubits which are not measured) at the end of a quantum circuit may be assumed to be measured \cite{Chuang}. Furthermore, the Toffoli gate in Fig.~\ref{eq05} can be decomposed into Hadamard, phase, CNOT, and $\pi/8$ gates \cite{Chuang}.

By further observing the circuit in Fig.~\ref{eq05}, we could eliminate the last CNOT and $X$ gates lying at the end without changing the output state, hence preserving the performance of this quantum adder and reducing the experimental error. So far, we have achieved decomposing our basis quantum adder $U$ into 11 CNOTs and 23 single qubit rotations (one Hadamard gate counts as two rotations: a $\pi/2$ rotation along the $y$-axis followed by a $\pi$ rotation along the $x$-axis), which in total add up to 34 quantum gates.

The fidelity of the quantum adder $\tilde{U}$ is defined as a function of the output state $\rho_{out}$ as
\begin{eqnarray}
&&F={\rm Tr}(\ket{\Psi_{id}}\bra{\Psi_{id}}\,\rho_{out}) \label{fid}, \\ \nonumber &&\rho_{out}={\rm Tr}_{12}(\tilde{U}\ket{\psi_1}\bra{\psi_1}\otimes\ket{\psi_2}\bra{\psi_2}\otimes\ket{0}\bra{0}\tilde{U}^\dag),
\end{eqnarray}
where the partial trace is taken over the first two qubits. We have plotted the fidelity of the basis quantum adder derived above in Fig.~\ref{gad}a. While showing a high theoretical fidelity, the experimental one is estimated by the gate errors reported by the Google labs group~\cite{superconducting}, which is about $1\%$ for a two-qubit controlled-Phase gate and $0.1\%$ for an arbitrary single-qubit gate. Recalling each CNOT gate can be realized by one controlled-Phase and two Hadamard gates, if the average theoretical fidelity is $F_{a}$, then an estimation of the experimental fidelity of the quantum adder is,
\begin{align}
F_{\rm exp}=F_{a}\times(0.999)^{N_s +2N_{\rm CNOT}}\times(0.99)^{N_{\rm CNOT}}.
\label{fexp}
\end{align}
Here, $N_s$ and $N_{\rm CNOT}$ stand for the number of single-qubit gates and the number of CNOT gates, respectively. After we take Eq.~\eqref{fexp} into account, the remaining experimental fidelity is about $80\%$, which is still high. We point out that, in order to implement the circuit of Fig.~\ref{eq05} with 11 CNOTs using a setup as the Google labs one~\cite{superconducting} with nearest neighbour coupling, a triangular 3-qubit geometry in the superconducting circuit may be straightforwardly employed. 

\begin{figure}[h]
{\includegraphics[width=0.5\textwidth]{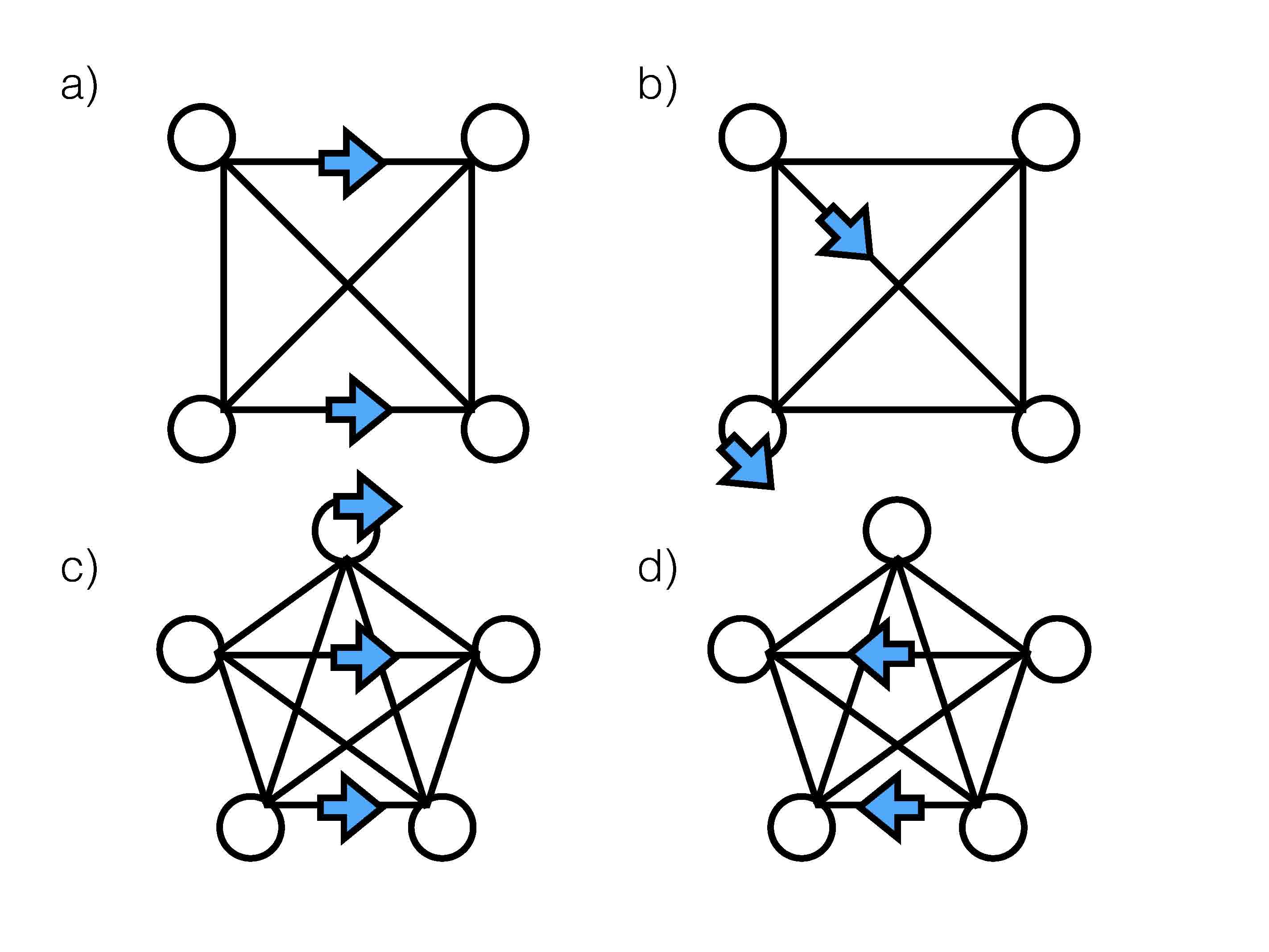}} 
\caption{{\bf Scheme of optimal grouping of $B_{ij}$ states.} (a) For even dimension $d$, tuples of $B_{ij}$ states are obtained by grouping the vector connecting vertices $i$ and $i+1$ with all the parallel vectors in the same direction. (b) The remaining $d$ tuples are obtained by grouping the monogon of each vertex with all the vectors perpendicular to the line that connects the vertex with its opposite one in a given direction. (c) For odd dimension $d$, tuples of $B_{ij}$ states are obtained by grouping the vector connecting vertices $i$ and $i+1$ with the monogon of the opposite vertex and all the diagonals parallel to the vector in the same direction. (d) The remaining $d$ tuples are obtained with similar procedure for the opposite direction but excluding the monogons of the vertices, because they were already grouped.}
\label{fig0}
\end{figure}

The basis quantum adder can be generalized to act on qudits of dimension $d$. The simplest expression consists in defining the adder $U$,  superposing the elements of the basis with a residual subspace exclusive of those particular elements, 
\begin{equation} 
U\ket{i}\ket{j}\ket{A}=\ket{i+j}\ket{B_{ij}}.
\end{equation}
Here, $\ket{i+j}$ represents $\frac{1}{\sqrt{(2+2 \delta_{ij})}} \left( \ket{i}+\ket{j} \right)$, and the ancillary state $\ket{A}$ has the same dimension as the input states, which enables that $\langle B_{\alpha\beta} |  B_{ij} \rangle=\delta_{\alpha i}\delta_{\beta j}$ is satisfied $\forall$ $i, j=0,...,d-1$. In order to reduce the resources and enhance the fidelity, we provide an alternative definition of the constituents of $U$ in which the number of residual states $B_{ij}$ is only $2d$ instead of $d^2$, which allows one to replace the $d$ dimensional ancillary state $\ket{A}$ with a qubit. This idea is supported by the fact that not all $B_{ij}$ need to be orthonormal for the unitarity conditions to be satisfied. The $B_{ij}$ can be combined in tuples of states that are represented with a single one, therefore reducing the dimension of the residual subspace. After analyzing this method for the low dimensional cases $d\le6$, we provide a discussion about its validity for any $d$.

\begin{figure*}[h]
{\includegraphics[width=\textwidth]{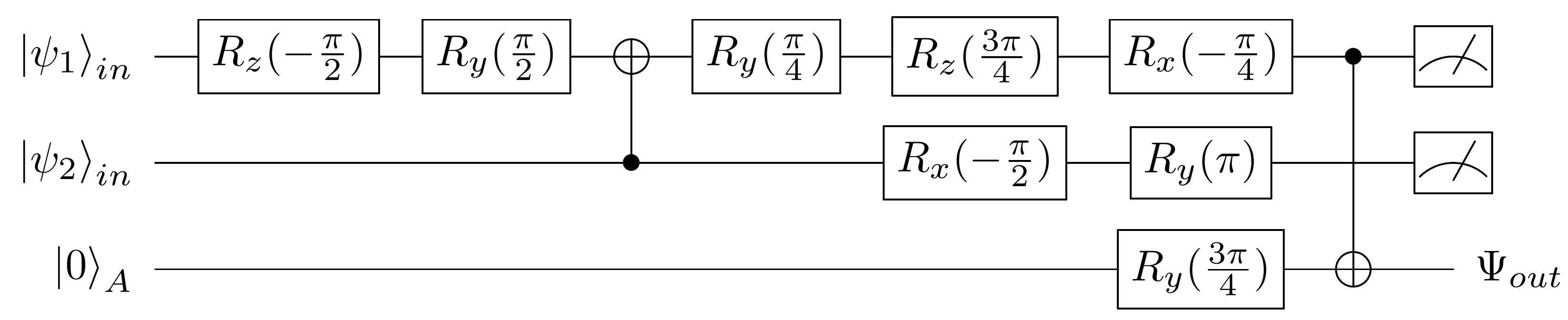}} 
\caption{{\bf Quantum circuit for a gate-limited quantum adder comprising only two CNOTs.} }
\label{g10}
\end{figure*}

The procedure is to count the number of tuples of $B_{ij}$ that do not need to be orthonormal, and contain the whole set of the $d^2$ $B_{ij}$ states. This $d^2$ is the total number of $B_{ij}$ if all of them were orthonormal. The unitarity condition implies that a set of orthonormal states is mapped into a set of orthonormal states, therefore $\ket{B_{\alpha\beta}}$ and $\ket{B_{ij}}$ only need to be orthonormal when any of the $\{\alpha, \beta\}$ coincides with any of the $\{i,j\}$. The reason is that $\langle \alpha + \beta | i + j \rangle = \frac{1}{2}\left( \langle \alpha | i\rangle +  \langle \alpha | j\rangle +   \langle \beta | i\rangle +  \langle \beta | j\rangle\right)$. Our argument is that the problem of finding the minimum number of tuples can be encoded in the structure of regular convex polygons of $d$ vertices. Each vector in a given direction between vertices $i$ and $j$ encodes a $B_{ij}$ element, while the opposite vector encodes the $B_{ji}$ for $i\neq j$. Additionally, monogons in each vertex encode $B_{ij}$ for $i=j$. Notice that the sum of the vertex monogons $d$ with twice the diagonals $d(d-3)$ and the sides $2d$ equals $d^2$, the total number of $B_{ij}$ if all were orthonormal. In the graphical analogy, the rule for obtaining tuples of $B_{ij}$ states that do not need to be orthonormal is to group the sides, diagonals or monogons that do not share any vertex. More precisely, we provide a method that guarantees that the number of tuples is $2d$. For even $d$, each of the $d$ tuples is obtained when grouping the vector $i,i+1$ with all the parallel diagonals and the vector in the opposite side and same direction. The remaining $d$ tuples are obtained when grouping the monogons in each vertex $i$ with the diagonals that are perpendicular to the diagonal that connects the vertex $i$ with its opposite vertex. For odd $d$, the $d$ tuples are obtained when grouping the vector $i,i+1$ with all the parallel diagonals and the monogon at the opposite vertex. The remaining $d$ tuples are obtained when grouping the same vector and diagonals in the opposite direction. See Fig. \ref{fig0} for a scheme of the analogy between $B_{ij}$ states and the regular convex polygons.            

Therefore, a set of $2d$ $\ket{B_{ij}}$ states is enough to satisfy the unitarity conditions, implying that only an ancillary with dimension $2$ is required. See, as an example, all the tuples for $d=4$ and $d=5$,
\begin{align*}
d=4:& \hspace{0.20cm} \{B_{01}, B_{32} \}, \{B_{12}, B_{03} \}, \{B_{23}, B_{10} \}, \{B_{30}, B_{21} \}, \\ &\{B_{00}, B_{13} \}, \{B_{11}, B_{20} \}, \{B_{22}, B_{31} \}, \{B_{33}, B_{02} \}. \\ 
\end{align*}
\begin{align*}
d=5:& \hspace{0.20cm} \{B_{01}, B_{42}, B_{33} \}, \{B_{12}, B_{03}, B_{44} \}, \{B_{23}, B_{14}, B_{00} \}, \\ &\{B_{34}, B_{20}, B_{11} \}, \{B_{40}, B_{31}, B_{22} \}, \{B_{10}, B_{24} \}, \\ &\{B_{21}, B_{30} \}, \{B_{32}, B_{41} \}, \{B_{43}, B_{02} \}, \{B_{04}, B_{13} \}.
\end{align*}

\begin{figure*}[t]
{\includegraphics[width=\textwidth]{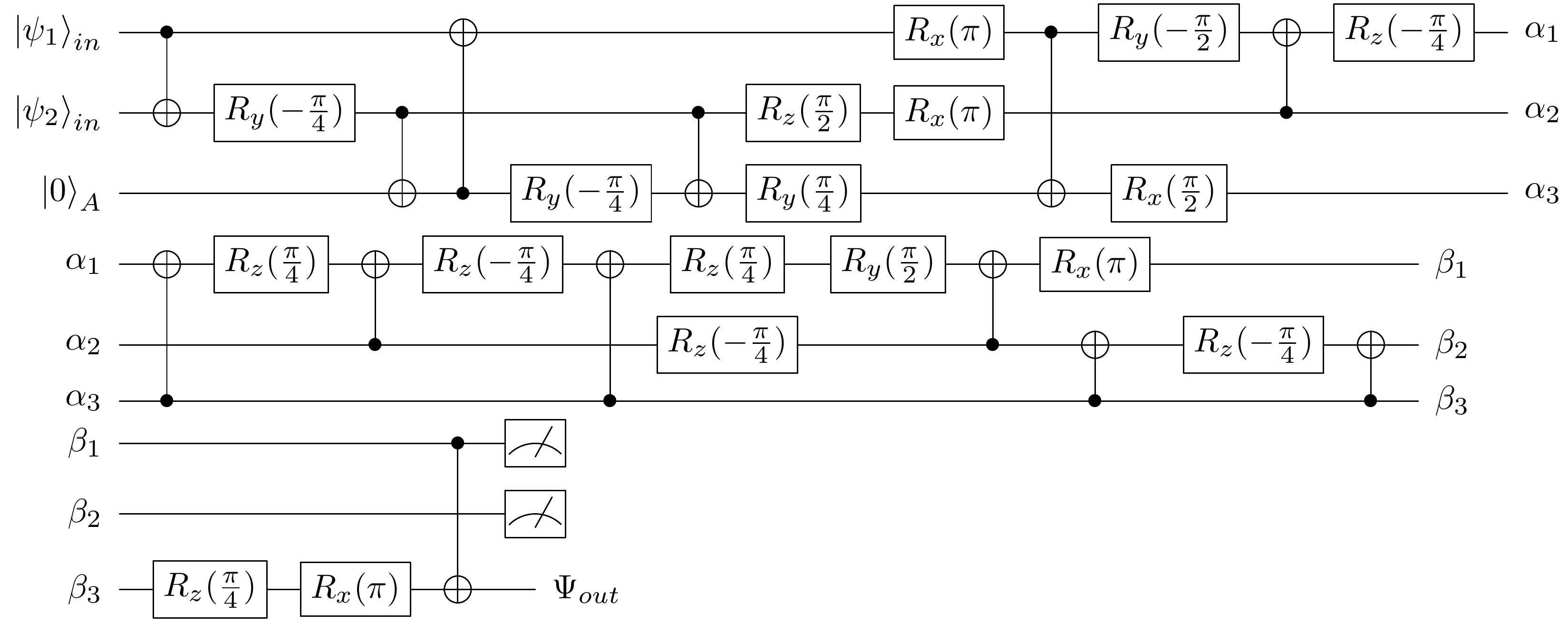}} 
\caption{{\bf Quantum circuit for a 40-gate approximate quantum adder.} }
\label{g31}
\end{figure*}

\begin{table*}
\begin{tabular}{cc|c c c c c c c c c|}
\cline{3-11}
& & \multicolumn{9}{ c| }{Newborn} \\ \cline{3-11}
& & 1 & 2 & 3 & 4 & 5 & 6 & 7 & 8 & 9  \\ \cline{1-11}
\multicolumn{1}{ |c  }{\multirow{4}{*}{Parent} } &
\multicolumn{1}{ |c| }{1} & p-2 & p-2 & p-1 & p-2 & p-1 & p-1 & 0 & 0 & 0     \\ \cline{2-11}
\multicolumn{1}{ |c  }{}                        &
\multicolumn{1}{ |c| }{2} & 2 & 2 & 1 & 0 & 0 & 0 & p-1 & p-1 & 0     \\ \cline{2-11}
\multicolumn{1}{ |c  }{}                        &
\multicolumn{1}{ |c| }{3} & 0 & 0 & 0 & 2 & 1 & 0 & 1 & 0 & p-1      \\ \cline{2-11}
\multicolumn{1}{ |c  }{}                        &
\multicolumn{1}{ |c| }{4} & 0 & 0 & 0 & 0 & 0 & 1 & 0 & 1 & 1       \\ \cline{1-11}
\end{tabular}
\caption{We show the amount of the total $p$ rows of each newborn individual that come from a given parent individual. The specific rows to change are randomly selected.}\label{Tablep}
\end{table*}

\begin{figure}[h]
{\includegraphics[width=0.5\textwidth]{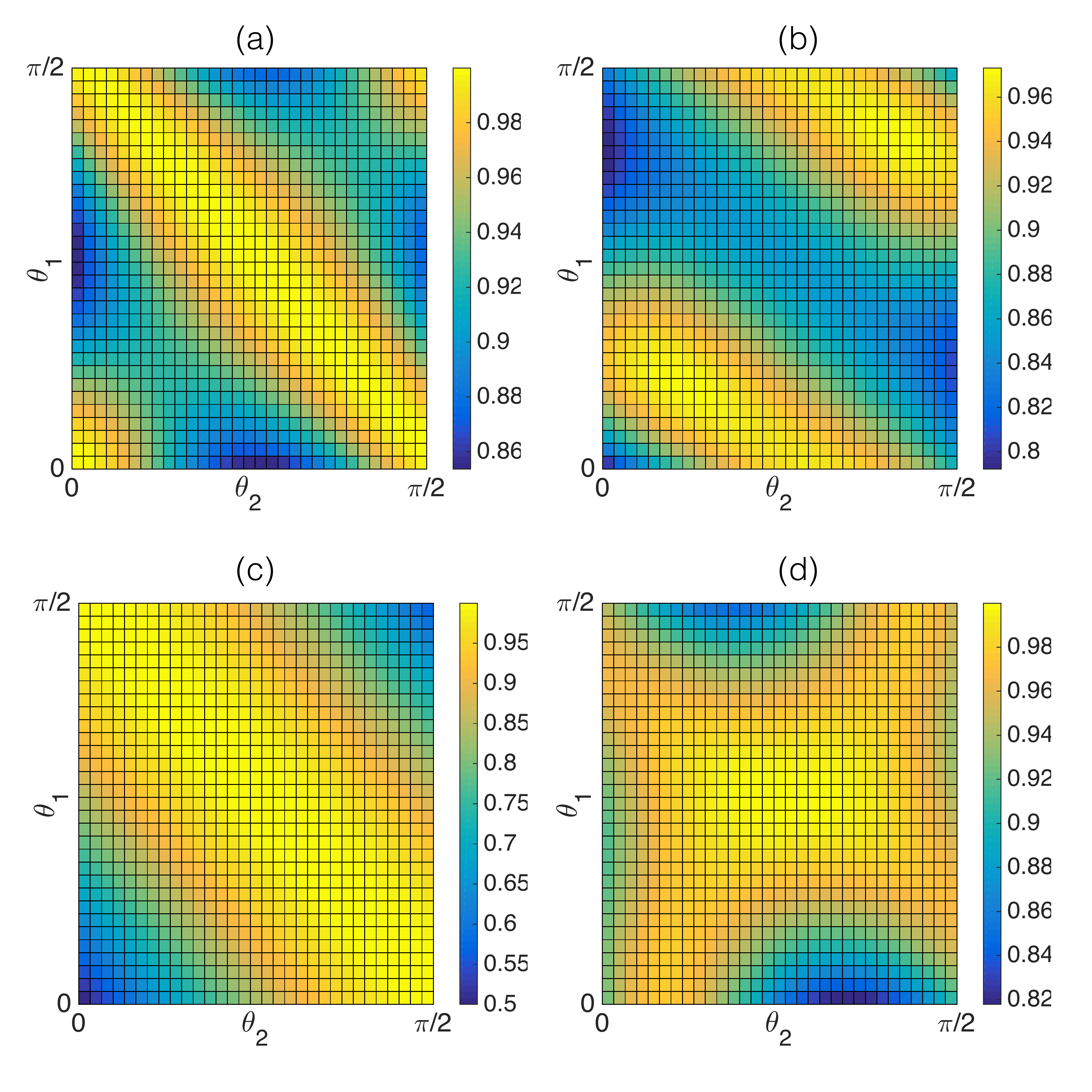}}
\caption{{\bf Fidelities of approximate quantum adders.} (a) Fidelity of the basis quantum adder vs $\theta_1,\theta_2$. The average fidelity of this region is $94.9\%$, while the lowest fidelity is $85.4\%$. (b) Fidelity of the gate-limited quantum adder vs $\theta_1$ and $\theta_2$. The average fidelity (theoretical) of this region is $90.0\%$, while the minimum one is $79.2\%$. (c) Fidelity of the trivial quantum adder given by the $\ket{+}$ state, with an average theoretical fidelity of $90.2\%$ and a minimum fidelity of $50\%$. (d) Fidelity of the 31-gate approximate quantum adder vs $\theta_1$ and $\theta_2$. The average theoretical fidelity of this region is $95.4\%$, while the minimum fidelity is $81.2\%$.}
\label{gad}
\end{figure}

\subsection{Genetic algorithms}

With the goal of improving the basis quantum adder, we have developed a program using genetic algorithms~\cite{ga5} to find the optimal protocols for the adding operation. The algorithm works due to the formalism derived to translate each quantum circuit diagram to a sequence of instructions, and the fidelity, to its analogous fitness function. The algorithm developed here, which was first introduced as a tool for optimizing digital quantum simulations~\cite{qga4}, has been adapted to account for gate decomposition problems in the superconducting quantum circuit platform.

Each cycle in the algorithm starts with four $p\times3$ matrices, representing four sequences of gates from a finite set of gates, where $p$ stands for the maximum number of gates allowed in the protocol which can be arbitrarily chosen. Each row in the matrices specifies a quantum gate from the set $S_g$,
\begin{align}
  S_g &=  \{R_x^{(i)}(\theta),\, R_y^{(i)}(\theta),\,R_z^{(i)}(\theta),\, U_{\rm CNOT}^{(i,j)}\} \\i,j&=\{1,2,3\};~ \theta=\{ \pi, \pi/2, \pi/4, -\pi/4, -\pi/2, -\pi\}. \nonumber
\end{align}
with 61 possibilities ($U_{\rm CNOT}^{(i,i)}=I$). The initial population can be either randomly or purposely chosen, depending on the convenience of introducing a previous solution. Firstly, the individuals have to be sorted according to their corresponding fidelity. Afterwards, the genetic algorithm hierarchically recombines the rows between different individuals, generating several new-born sequences with the same number of rows. Nine new individuals are created in this stage, from which six, five, four and three contain information of the first, second, third and fourth individuals respectively. These numbers arise from the recombination mechanism, according to which, each new individual copies most of its genetic code from a dominant individual and only a small fraction from a second recessive one. The role is determined by the previously mentioned ordering according to their fidelity. See Table~\ref{Tablep} for a schematic representation of the recombination.

The next step in the algorithm is the mutation stage. In this, a row of the newborn individual is exchanged by a randomly generated one if a random number exceeds the mutation threshold.  After the mutations, all the newly generated and the original input sequences will be sorted according to their fidelity given by Eq.~(\ref{fid}). Finally, the highest four sequences will be selected and kickstart the forthcoming cycle as the initial inputs. One can specify the total number of generations and maximum number of gates in the fidelity or circuit optimization. The more rows we allow for our protocol, the better it can approximate a potential optimal quantum adder $U$, since the versatility for realizing an arbitrary unitary matrix gets improved. However, it will be harder for the protocol to be carried out in a laboratory due to the increasing complexity. Hence, we have to make a compromise and set a limit of $p$ according to physical conditions allowed in each particular lab.

An important remark to mention is that the fidelity is calculated on pairs of states. Therefore, in order to evaluate the gate sequence on the complete Hilbert space we have discretized it and employed either the minimal or the average fidelities. Additionally, notice that the parameters encoding the action of the algorithm, i.e., the recombination fraction and the mutation mechanism and threshold, may be tuned for balancing the behavior of the search process between converging to a local minima and exploring the complete space of solutions.

\begin{table*}[h!!]
\begin{tabular}{||c c c c c c c||}
\hline
\multicolumn{7}{|c|}{IBM Quantum Experience}\\
\hline
$\theta_1,\theta_2$ & $0,0$ & $\pi/2,\pi/2$ & $0,\pi/2$& $\pi/2,0$ & $\pi/4,\pi/4$ & $\pi/8,\pi/8$\\ \hline
Experimental results & 0.815  & 0.749 & 0.873 & 0.853 & 0.839 & 0.935\\ \hline
Classical ideal simulation & 0.802 & 0.802 & 0.854 & 0.854 & 0.854 & 0.963 \\
\hline
\end{tabular}
\caption{Fidelity $F={\rm Tr}(\ket{\Psi_{id}}\bra{\Psi_{id}}\,\rho_{out})$ of the outcome $\rho_{out}$ for the gate-limited quantum adder of Fig.~\ref{g10} with respect to the ideal sum $\ket{\Psi_{id}}$. We include the experimental results employing IBM Quantum Experience 5-transmon device, as well as the classical ideal simulation. Each experimental value involves 1024 measurement shots.}\label{TableIBM}
\end{table*}

\subsection{Quantum adders found by genetic algorithms}

By setting the maximum number of gates to 20, we have found a gate-limited quantum adder consisting of only two CNOTs having an average theoretical fidelity of $90\%$ and a minimum of $79.2\%$. (see circuit in Fig.~\ref{g10} and Fig.~\ref{gad}b). Although its theoretical fidelity is lower than the one of the basis adder ($94.9\%$), its implementation fidelity is actually the highest one, about $87\%$, if implemented in superconducting circuit platforms~\cite{superconducting}. An interesting point to highlight here is that this quantum adder has nearly the same average fidelity as the one given by a plus state, $|+\rangle$, in the output of the adder (see Fig.~\ref{gad}c). The difference is that $|+\rangle$, which is trivial because it does not depend on the inputs, has a lower minimal fidelity of $50\%$ and an average theoretical fidelity of $90.2\%$. Nevertheless, this trivial quantum adder establishes the lower limit of the average fidelity for the quantum adder to be considered interesting in the region we are confined to.

If we allow for 40 gates, the GA achieves an approximate quantum adder with an average theoretical fidelity above $95\%$ (see Fig.~\ref{gad}d and circuit in Fig.~\ref{g31}). This quantum adder contains 31 gates overall, 13 of which are two-qubit CNOT gates. The expected experimental fidelity of this 31-gate quantum adder is roughly the same as the basis adder which is about 80$\%$.

It should be noticed that this quantum adder and the one we found previously with 10 gates defined in the circuit in Fig.~\ref{g10} are not commutative quantum adders, i.e., the quantum adding machine $\mathcal{M}$ defined by them does not satisfy
\begin{equation}
\mathcal{M}(\psi_1,\,\psi_2) = \mathcal{M}(\psi_2,\,\psi_1)
\end{equation}
for arbitrary input states $\psi_1$ and $\psi_2$. The main reason is that the GA does not select the gate sequence according to the commutativity of the resulting unitary, but according to the average fidelity of the quantum adder. Another result to highlight is the absence of a high-fidelity and universal quantum adder. The only result obtained so far in this respect is a fixed quantum state, with an overall fidelity of $50\%$ and independent of the inputs, which is perpendicular to the region in which the quantum adder is defined. This result coincides with the classical limit of randomly choosing a qubit state. 

\begin{figure*}[h]
{\includegraphics[width=\textwidth]{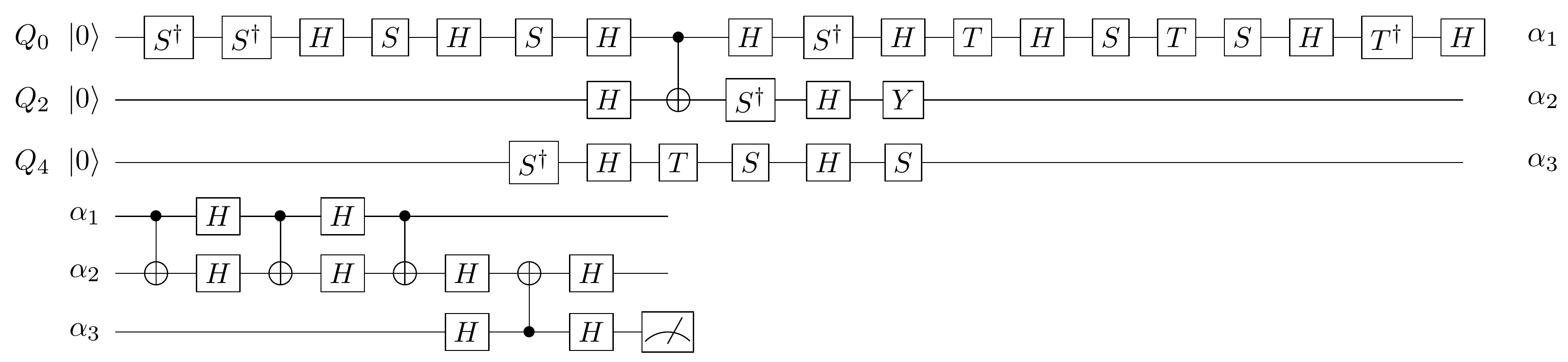}} 
\caption{{\bf Quantum circuit for the gate-limited quantum adder of Fig.~\ref{g10} recast in terms of the 
Clifford set available on IBM Quantum Experience.} }
\label{ibm}
\end{figure*}

\subsection{Experimental realization on IBM Quantum Experience}

We have experimentally realized the gate-limited quantum adder in the 5-transmon quantum computer provided by the facilities of IBM Quantum Experience. We have considered three qubits for this purpose, two of them encoding the initial addend states and the third one encoding the ancilla. We have rewritten the quantum circuit in Fig.~\ref{g10} in terms of the Clifford set available in IBM Quantum Experience, as seen in Fig~\ref{ibm}. Here, the quantum adder in Fig.~\ref{g10} is rewritten in the language of IBM Quantum Experience. Qubits $Q_0$ and $Q_2$ denote the two input states to be added, and $Q_4$ is the ancilla qubit onto which the approximate sum is given as output. The case shown is for $\{\theta_1,\theta_2\}=\{0,0\}$. The boxes denote Clifford group gates, the last one being a Z basis measurement.

We show the experimental results in Table~\ref{TableIBM}. We include the measurements in the IBM 5 transmon quantum computer and the ideal theoretical predictions. One can appreciate that the agreement between theory and experiment is significant, with a deviation smaller than 10\% in all cases. Therefore, this quantum platform allows for a high-fidelity quantum adder with current technology and, as gate fidelities improve, the accumulated gate error will be further reduced.

\section{Discussion}
We have studied the existence of an optimal and consistent approximate quantum adder with the support of genetic algorithm techniques in a specific region of the Hilbert space. Explicit protocols of three approximate quantum adders have been studied, while considering a suitable balance between average fidelity, number of gates, and experimental errors. The technique for approximating the quantum adder with genetic algorithms could be useful to extend current results to higher dimensions of ancillary and input quantum states. We have also experimentally implemented the proposed quantum adder on the quantum computer provided by IBM Quantum Experience, thus demonstrating its feasibility. Quantum adders have already been proven to be useful as a building block for the development of quantum algorithms \cite{qa5}. Therefore, the study of approximate and efficient quantum adders represents a fundamental theoretical challenge and a route towards improved quantum protocols.

\section{Acknowledgments}
We acknowledge use of the IBM Quantum Experience for this work. The views expressed are those of the authors and do not reflect the official policy or position of IBM or the IBM Quantum Experience team. The authors acknowledge support from Spanish MINECO/FEDER FIS2015-69983-P, Ram\'{o}n y Cajal Grant RYC-2012-11391, UPV/EHU UFI 11/55, Basque Government BFI-2012-322 and IT986-16.

\end{document}